# Epitaxial Graphene on Silicon toward Graphene-Silicon Fusion Electronics

*Hirokazu Fukidome[1,2], Ryota Takahashi[1], Yu Miyamoto[1], Hiroyuki Handa[1,2], Hyun-Chul Kang[1,2], Hiromi Karasawa[1], Tetsuya Suemitsu[1,2], Taiichi Otsuji[1,2] Akitaka Yoshigoe[3], Yuden Teraoka[3], Maki Suemitsu[1,2]

[1] *Research Institute of Electrical Communication, Tohoku University, Sendai 980-8577, Japan*

[2] *CREST, Japan Science and Technology Agency, Chiyoda, Tokyo 102-0075, Japan*

[3] *Japan Atomic Energy Agency, 1-1-1 Kouto, Sayo-cho, Sayo-gun, Hyogo 679-5148, Japan*

**Graphene is a promising contender to succeed the silicon's throne in electronics (beyond CMOS)[1]. To this goal, large-scale epitaxial growth of graphene on substrates should be developed[2-5]. Among various methods along this line, epitaxial growth of graphene on SiC substrates by thermal decomposition of surface layers has proved itself quite satisfactory both in quality[2] and in process reliability. Even modulation of structural and hence electronic properties of graphene is possible by tuning the graphene/SiC interface structure[6,7]. The challenges for this *graphene-on-SiC* technology, however, are the abdication of the well-established Si technologies and the high production cost of the SiC bulk crystals[1]. Here, we demonstrate that formation of epitaxial graphene on silicon substrate is possible, by graphitizing epitaxial SiC thin films formed on silicon substrates[8-10]. This *graphene-on-silicon* (GOS) method enables us to form a large-area film of well-ordered $sp^2$ carbon networks on Si substrates and to fabricate electronic**

**devices based on the structure.**

The GOS process starts with a 3C-SiC growth on Si substrates. Our previous work has proven that use of monomethylsilane (MMS) as a single source for SiC epitaxy, instead of monosilane-propane two source system, is quite successful in lowering the growth temperature without any serious degradation of the film[11]. Figure 1a shows the C1$s$ core-level spectrum before (orange) and after (blue) the graphitization (1523 K, 30 min anneal *in vacuo*). Before the graphitization, the surface consists predominantly of SiC. This is in good agreement with the low-energy electron diffraction (LEED) pattern (Fig. 1b) obtained from the same surface, showing hexagonal (1×1) and Si-rich ($\sqrt{3}\times\sqrt{3}$)$R$30º spots of a 3C-SiC(111) thin film on Si(111) (Fig. 1b). The thermal graphitization makes the surface predominately covered with $sp^2$ carbon atoms in Fig. 1a. The LEED observation now presents a hexagonal (1×1) pattern, which is rotated by 30 ° with respect to that of 3C-SiC(111). This rotation accords with the change in the surface atomistic structure from 3C-SiC(111) to graphene[12]. The changes in LEED and C1$s$ core level spectroscopy altogether reveal onset of real epitaxy of graphene on SiC thin films, which proceeds as in the same manner as what happens during epitaxy of graphene on SiC(0001) bulk crystals[6].

The viability of GOS for device applications is demonstrated in Fig. 2. First of all, uniformity of the GOS material, crucial for a large-scale production, is shown in Fig. 2a and 2b. The spectra of Raman microspectroscopy (Fig. 2b), taken at points indicated as red dots in Fig. 2a, all present the principal vibration modes of graphene, G (~1590 cm$^{-1}$) and G' (~2720 cm$^{-1}$) bands[13,14] with similar intensities, demonstrating a good uniformity of the present graphene within the 3cm$^2$ area. Thanks to this uniformity, GOS films can be processed to fabricate field-effect transistors (FET) using a standard lithographic techniques for Si. The device structure of a GOS-based FET (GOS-FET), thus made, is schematically shown in Fig. 2c. Figure 2d shows a microscope photograph of several transistors fabricated on the GOS substrate. A typical drain current ($I_D$)-drain voltage ($V_{DS}$) characteristics are shown in Fig. 2e. The $I_D$ values are one order of magnitude larger than those obtained without forming graphene layers. The effective carrier mobility is estimated to be more than 2000 cm$^2$/(V.s) for the back-gate FET sample on which no SiN layer was deposited[15]. However, the currents shown in Fig. 2e are much smaller than that estimated from such a large mobility. This suggests decrease in the mobility presumably due to the influence of the SiN gate stack and the plasma damage caused by plasma-enhanced chemical vapor deposition of SiN. The linear region transconductance ($G_m$) measured at $V_{DS}$ = 10 mV is shown in Fig. 2f. In this figure six devices distributed in a 4 mm × 4 mm chip are tested. The obtained $G_m$ values for these devices reveal quite similar behavior in their gate voltage dependency. This is again due to the good uniformly of GOS, proving its high potential as a material for device integration.

Although much is shared between the properties of GOS and epitaxial graphene on SiC, one can suggest at least two points as uniqueness of GOS. The first one is the GOS's degree of freedom in the choice of crystallographic orientation of the 3C-SiC thin film, which is accomplished by changing the orientation of the Si substrate. In Fig. 3a, we compare the C1$s$ core level spectra of thermally graphitized 3C-SiC(111)/Si(111), 3C-SiC(100)/Si(100) and 3C-SiC(110)/Si(110). In all the spectra, the peak due to $sp^2$ carbon atoms as well as that from SiC bulk appear. Raman spectra of GOS on these three configurations are displayed in Fig. 3b. In all the spectra, the principal vibrational features of graphene, G and G' bands, appear. It is thus concluded that epitaxial graphene can be grown on three major low-index surfaces of 3C-SiC thin film and, equivalently, on three major low-index Si substrates. On bulk SiC crystals, on the other hand, graphene epitaxy has been so far limited to hexagonal faces, such as SiC($0001$) and SiC($000\bar{1}$)[3-7]. Although one cannot deny the possibility of graphene epitaxy on other crystallographic orientations than c-axis faces of 6H- or 4H-SiC crystals, this would surely boost the price of the EG process that is already high. The GOS growth on three major low-index Si surfaces will fuel the feasibility of GOS process in the industry.

The second unique point of GOS is the tunability of the interface structure between graphene and SiC. This is accomplished, again, by the choice of the surface orientation of the 3C-SiC thin film. In the C1$s$ core-level spectra in Fig. 3a, a peak is found at the higher binding energy side of the main $sp^2$ peak on SiC(111)/Si(111). This peak is related to the interfacial layer that experiences a charge transfer in the incommensurate system of graphene and SiC(111) system[16]. Previous studies on epitaxial graphene on hexagonal SiC have clarified that the graphene layers on Si-terminated SiC(0001) surface is mainly Bernal-stacked and is accompanied by an interfacial layer[3,4,6,7]. On C-terminated SiC($000\bar{1}$) surface, on the other hand, graphene is non-Bernal-stacked without the interfacial layers. Applying this knowledge to GOS, it is suggested that the graphene on 3C-SiC(111)/Si(111) surface is Bernal-stacked, while that on 3C-SiC(100)/Si(100) or 3C-SiC(110)/Si(110) surfaces is non-Bernal-stacked.

This hypothesis is actually supported by the lineshape analysis of the G' bands in the Raman spectra in Fig. 3b. Only the graphene on 3C-SiC(111)/Si(111) shows the G' band that involves multiple components. Among the components are the one related to 2D graphite (2717 cm$^{-1}$) and to Bernal-stacking graphene (at ~ 2620 cm$^{-1}$ and ~ 2750

cm$^{-1}$)$^{13,14}$. The splitting of the G' band due to Bernal stacking is also observed for epitaxial graphene on Si-face SiC($0001$) bulk crystals$^{17}$. In sharp contrast, the G' bands of graphene on 3C-SiC(100)/Si(100) and 3C-SiC(110)/Si(110) surfaces can be expressed by a single component (~ 2725 cm$^{-1}$), similar to epitaxial graphene on C-face SiC($000\bar{1}$). In the latter, non-Bernal stacking of epitaxial graphene is caused by the absence of the interfacial layer$^{18}$. This tunability of the graphene stacking as well as of the interfacial structure is definitely a benefit of the GOS structure.

These two points mentioned above may give unprecedented merits to the GOS technology. This is especially true when one considers its fusion with the advanced 3D-structuring of Si devices (3D-GOS), which is easily accessed by the state-of-the-art Si nanopatterning technology. Even a 3D-GOS FET can be imagined, in which a semiconductive graphene portion on 3C-SiC(111)/Si(111) is used as the channel while the surrounding metallic graphene portions on 3C-SiC(110)/Si(100) or 3C-SiC(110)/Si(110) microfacets are used as the source/drain electrodes or interconnects.

In summary, we have demonstrated that epitaxy of graphene on silicon substrates in large area is viable toward realistic device applications of graphene into graphene-silicon fusion electronics. The future of GOS is not limited to its fusion with Si technology, but includes novel technologies featuring its unique tunability over the electronic properties of the graphene. By forming GOS films on 3D-structured Si substrates (3D-GOS), one may fabricate various components using solely the GOS structures.

**Methods**

Substrates were cut from Si(100), Si(110) and Si(111) wafers into 7 × 40 mm$^2$ pieces. The substrates are Boron-doped (1-10 Ω cm). The epitaxy of GOS consists of two steps. The first step is the epitaxial formation of 3C-SiC thin films on silicon substrates by gas-source molecular-beam epitaxy (GSMBE) using monomethylsilane[8-11]. The crystallographic orientations of the 3C-SiC thin films used in this study are 3C-SiC(110)/Si(110), 3C-SiC(100)/Si(100) and 3C-SiC(111)/Si(111). The second step is graphitization of the SiC thin films by annealing the films in vacuo with a resistive heating at 1523 K for 30 min[8-10]. The crystal structure of the epitaxial films of SiC and graphene was in-situ monitored by a LEED apparatus with a base pressure of ~ 10$^{-8}$ Pa. C 1*s* core level photoelectron spectroscopy was carried out at the surface-chemistry endstation of BL23SU at SPring-8 with a base pressure of ~ 10$^{-8}$ Pa. The formation of graphene films was examined in air by Raman microspectroscopy (Renishaw)[8-10].

For the fabrication of GOS-FET[15], the ohmic electrodes are defined by the lift-off process with Ti/Au. The device isolation is carried out by oxygen plasma etching to remove the graphene out of the device area. As the gate stack, 200-nm thick SiN is deposited by plasma-enhanced chemical vapor deposition (PECVD). This is followed by the gate metallization with Ti/Au. The probing pads are connected to the ohmic electrodes via holes through the gate stack. The gate length is 10 μm and the channel width is 20 μm. Standard optical lithography with a mask aligner is used for all process steps.

**Legends**

**Figure 1 | Atomic and electronic structure of GOS**. **a**, Comparison of C1s core-level spectra of epitaxial 3C-SiC(111) thin films on Si(111) before (blue) and after (red) thermal graphitization at 1523 K for 30 min. **b**, LEED pattern at 53.7 eV showing the diffraction spots due to $(1\times1)$ (white arrows) and $(\sqrt{3}\times\sqrt{3})$ (blue arrows) the 3C-SiC(111) thin film. **c**, LEED pattern at 62 eV showing the diffraction spots due to the graphene lattice (red arrows).

**Figure 2 | Macroscopic homogeneity and characteristics of GOS-FET**. **a**, Optical micrographs of epitaxial graphene on Si(110). **b**, Spatially resolved Raman spectra of GOS taken at the positions indicated in **a**. **c**, Schematics of top-gate GOS-FET where the epitaxial SiC thin film and the Si substrate are used as the insulator and the gate, resopectively. **d**, The optical micrograph of GOS –FET fabricated by the standard lithography. **e**, Typical $I_{DS}$ versus $V_{DS}$ characteristics at different $V_G$ of GOS-FET. **f**, Linear region transconductance $|G_m|$ versus gate voltage of six GOS-FET devices (#1-#6) distributed in 4 mm x 4 mm chip area.

**Figure 3 | Surface and interface structures of epitaxial graphene on Si substrates**. **a**, Comparison of C1s core-level spectra of epitaxial graphene on Si(111) (red), Si(100) (blue) and Si(110) (black). The spectra of epitaxial graphene on Si(100) and Si(110) contains two contributions from the SiC thin film (marked SiC) and the graphene layer (marked G). On the other hand, the spectra of epitaxial graphene on Si(111) contains three components from the SiC thin film (marked SiC) and the graphene layer (marked

G), and the interfacial layer (marked I). **b**, Raman spectra of epitaxial graphene on Si(111), Si(100) and Si(110) substrates. The G' bands of epitaxial graphene on Si(100) and Si(110) is decomposed into a single component from non-Bernal stacked graphene layers. On the other hand, the G' band of epitaxial graphene on Si(111) is decomposed into three components owing to the Bernal stacking of graphene layers.

Figure 1

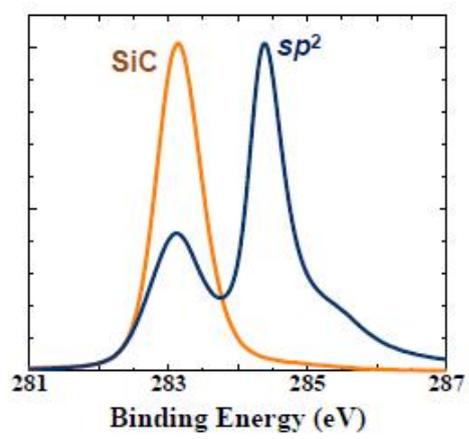 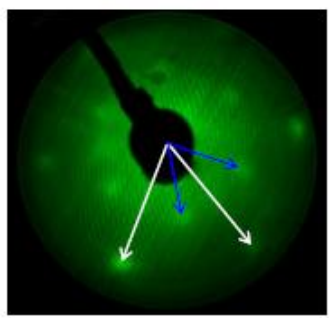 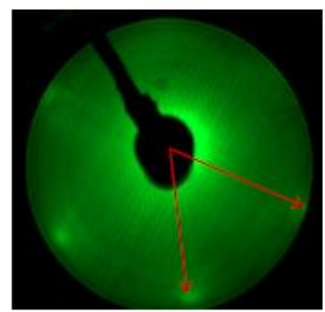

(a)          (b)          (c)

Fig. 2

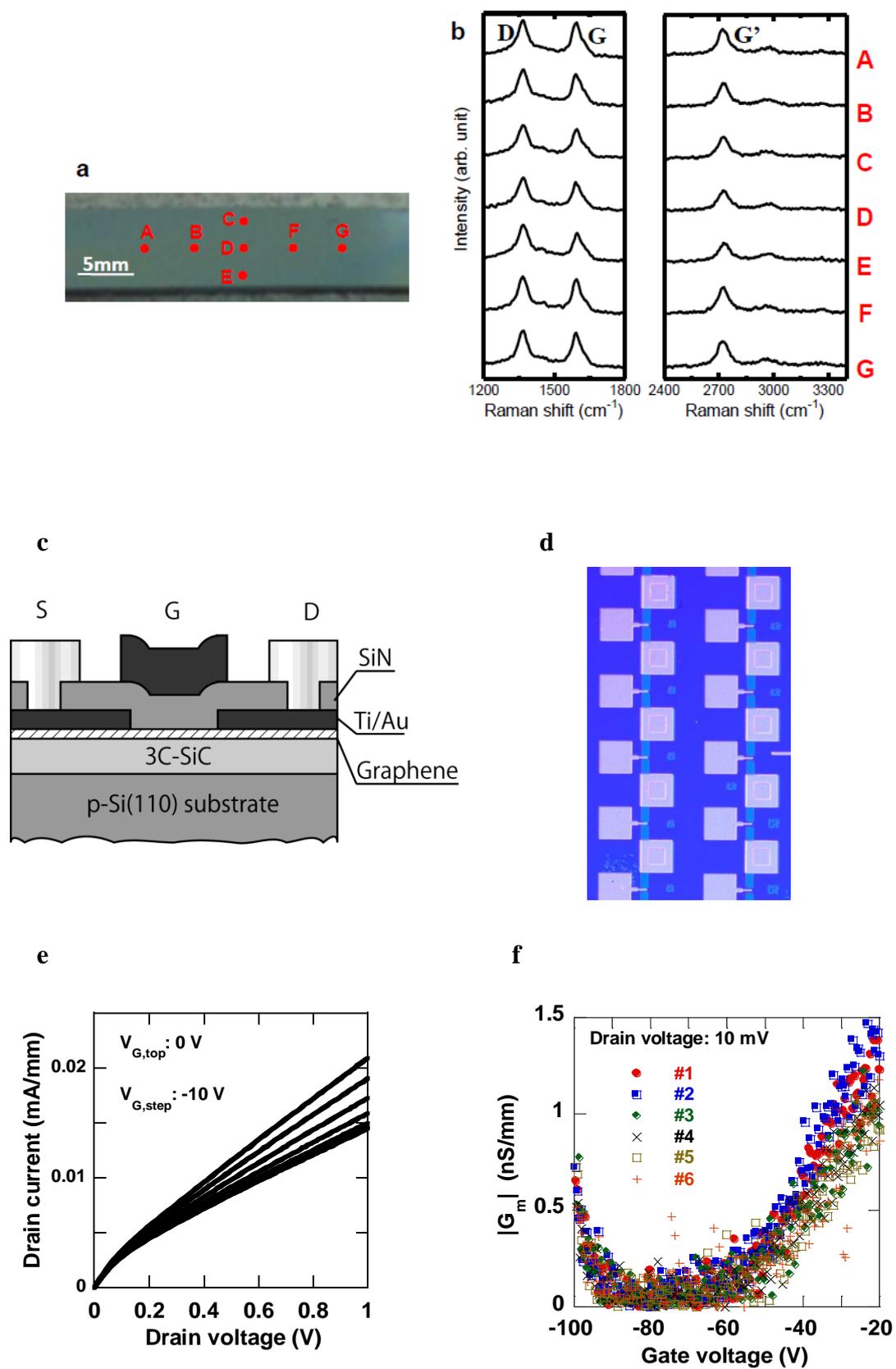

Fig. 3

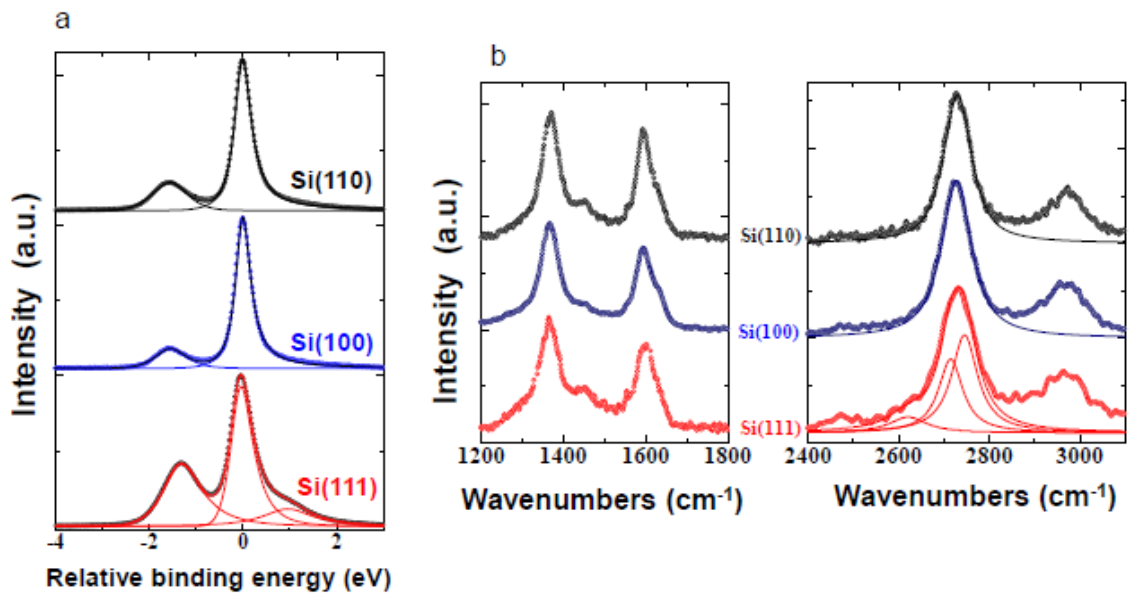